\documentstyle[preprint,aps]{revtex}
\begin{document}
\tightenlines
\title{Irradiation of benzene molecules by ion-induced and 
light-induced intense fields} 
\author{D. Mathur}
\address{Tata Institute of Fundamental Research, Homi Bhabha
Road, Mumbai 400 005, India}
\date{\today}
\maketitle
\begin{abstract}
Benzene, with its sea of delocalized $\pi$-electrons in the valence orbitals, 
is identified as an example of a class of molecules that enable establishment 
of the correspondence between intense ion-induced and laser-light-induced fields 
in experiments that probe ionization dynamics in temporal regimes 
spanning the attosecond and picosecond ranges.   
\end{abstract}

Matter is inherently unstable when exposed to electric fields whose magnitudes 
approximate interatomic Coulombic fields. Studies of the response of matter to 
very intense fluxes of such fields address fundamental issues concerning the 
physics of systems driven strongly away from equilibrium. There are two approaches 
that enable terrestrial access to such fields, and their use in probing the ionization 
(and dissociation) dynamics of atoms (and molecules): use of fast, highly-charged 
beams of heavy ions on the one hand, and focused, intense laser light on the other. 
The equivalence of the two approaches has only been explored in somewhat 
desultory fashion. The interaction of atoms and molecules with intense fields is 
clearly a complex nonperturbative, dynamical problem and prospects of rigorous 
theoretical treatment remain remote. Experimental information, albeit of a 
morphologial nature, is, therefore, of considerable importance and interest. 
Specifically, it is necessary to identify a class of target species whose properties 
make them suitable for studies aimed at disentangling the important time-dependent 
aspects of the field-target interaction from those that are 'static' or 
structure-dependent, so that the prospects of gaining qualitative insights are 
improved. The time-dependent aspect of the field-target interaction is especially 
of relevance in the case of molecules. Contemporary laser technology makes 
available intense light fields in pulses of picosecond ($ps$) and femtosecond 
($fs$) duration. For molecules, these time scales are of significance in 
relation to typical rotational and vibrational time periods (tens of $ps$
and tens of $fs$, respectively). In the case of ion-induced fields, 
however, a beam of highly-charged ions traverses a distance of a few {\AA} in only 
a few tens of attoseconds $as$, thereby opening prospects of studies of 
molecular dynamics 
on times scales in which even vibrational motions can be regarded as ``frozen". 
Moreover, these ultrashort times imply enormous uncertainties in energies, thereby 
opening new avenues of molecular dynamics studies that fall well outside the 
ambit of conventional molecular quantum mechanics. We report in this Letter results 
of a study that suggests that benzene, with its sea of delocalized $\pi$-electrons 
in the valence orbitals, constitutes a molecule that enables establishment of the 
correspondence between intense ion-induced and laser-light-induced fields in 
experimental probes of the ionization dynamics in temporal regimes that  
span the $as$ and $ps$ ranges. 

The correspondence between ion-induced and light-induced fields can most usefully 
be established by considering three temporal regimes. In the case of relativistic 
ion beams (whose energy$\sim$GeV/nucleon, $\beta$=$v/c\sim$1) colliding with atoms 
or molecules, the Weizs\"acker-Williams equivalent photon model \cite{ww} transforms 
the Coulombic field in the projectile's rest frame to the target's rest frame, and 
equates the effect of the field to two, orthogonally-directed photon pulses. 
Application of such equivalent-photon pulses to atomic ionization have 
recently attracted attention \cite{moshammer}. At the opposite temporal extreme, 
purely electrostatic considerations use Coulomb's law; for instance, in 
the case of Si$^{3+}$ projectiles, a target atom experiences a 
field of magnitude $\sim$5 V ${\AA}^{-1}$ ($\sim$0.1 a.u.) at a distance of 3 {\AA}. 
In the intermediate regime, one that is experimentally most accessible using 
highly-charged ion beams of MeV energy, the effective ion-induced electric field 
experienced by a target may be deduced from the Poynting vector: $I_{eff}(t) = {1\over{\mu_0}}|\vec{E}(t)\times \vec{B}(t)|$. Following the pioneering work of 
Rhodes and coworkers in this regime \cite{rhodes}, recent work has focused 
on observing similarities and differences between ion-induced and 
laser-induced ionization patterns in molecules like water and chloromethanes 
of different symmetries \cite{nsc}. For H$_2$O, 
the morphology of the dissociative ionization pattern was found to be grossly 
different in the two cases. Differences in the case of the 
chloromethanes were much less, but were still significant. These differences may be 
rationalized either by noting that the directional properties of the applied field 
are different in the two cases, or by invoking the different temporal profiles of 
the applied fields. Although the magnitudes of the ion-induced and light-induced 
fields were almost identical in these experiments, the directional properties of 
the two types of field were different in the following sense: the laser light was 
linearly-polarized and so, the direction of the light-induced field 
was constant in the course of the interaction, unlike the time-dependence that is 
intrinsic to the direction of the ion-induced field. There is some evidence that, 
in the case of laser-molecule interactions, the directional properties of the 
applied field can influence molecular dissociation pathways \cite{ch4}. Other 
factors, such as indirect ionization events (involving intermediate electron 
capture and loss) in the case of the ion-impact experiments, can be 
discounted by judicious choice of collision energy (and, as indicated below, by 
supplementary measurements made, under identical operating conditions, on 
well-understood collision systems involving rare-gas targets). 

Before presenting results that demonstrate the good correspondence between 
ion-induced and light-induced ionization patterns of $C_6H_6$, it is important to 
address other facets of the equivalence we explore, namely the 
peak field intensity and the temporal/spatial features for the two 
types of fields. Our ion-impact experiments used fast 
beams of mass-selected, highly-charged ions (Si$^{q+}$, $q$=3,8 at an energy of 
50-100 MeV) that were obtained from a tandem accelerator. Slow recoil ions 
resulting from large impact-parameter interactions between the projectile ions 
and $C_6H_6$ vapor were extracted, with unit collection efficiency, into a linear 
time-of-flight (TOF) spectrometer located in a direction that was orthogonal to 
the incident ion beam. The methodology has been described previously  
\cite{nsc}. Experiments were also performed with F$^{7+}$ projectiles at an energy 
of 110 MeV, and in these the ion analysis and detection was with 
quadrupole mass filter (all other experimental 
features being similar). In the laser-based experiments, light pulses (of 
35 ps and 100 fs duration), from high-intensity Nd:YAG and Ti:sapphire lasers,  
were focused by a 
biconvex lens of 10 cm focal length such that peak intensities within the focal 
volume were in the range 10$^{12}$-10$^{16}$ W cm$^{-2}$. Ions produced in the focal 
volume were extracted, again at 90$^\circ$ to the incident laser beam, and 
analyzed by a 2-field TOF set-up.     

One important facet of the comparison of ion-induced and light-induced 
ionization dynamics is the quantification of the peak magnitude of the applied 
field. A limiting value of the impact parameter, $b$, has to be deduced for each 
ion-collision system. However, for molecular targets, $b$ remains a 
somewhat elusive parameter. The range of $b$-values that come into play 
manifests itself in the mean recoil energy ($E_r$) that is imparted to each 
molecular ion that is created in the interaction. This, in turn, is reflected 
in the temporal width of molecualar ion peaks in the measured 
TOF spectrum. In the case of $C_6H_6^+$ ions formed in our experiments with 
Si$^{q+}$ ($q$=3,8) projectiles, we 
measured $E_r$ to lie in the range 30-40 meV. There is an established method 
\cite{tonuma}, based on classical trajectory Monte-Carlo techniques 
\cite{schlachter}, that enables deductions to be made of the impact 
parameter dependence of multiple ionization probabilities in fast-ion collisions. 
A value of 3 {\AA} was deduced as the lower limit for $b$ in our 
experiments. Collisions that occur at smaller $b$-values give rise to recoil 
energies far in excess of $E_r$, and these are discriminated against by the 
angular resolution of the spectrometers used by us.
We confirmed the veracity of our deductions of $b$ by determining total cross 
sections for formation of low-energy Ar$^{q+}$ recoils ($q$=1-10) 
in the same apparatus. Deduction of $b$ can also be complicated by electron 
capture and loss processes that might influence the formation of recoils in 
charge state $q>$1. However, it was confirmed that for a range of 
ions (namely, Si$^{q+}$, $q$=3-12), direct ionization dominated 
the overall dynamics. By way of example, we cite the measured the
total cross section 
for Ar$^{4+}$ formation in Si$^{10+}$-Ar collisions to be 8$\times$10$^{-17}$ 
cm$^2$; corresponding cross sections for Ar$^{4+}$ formation accompanied by 
1-electron capture and loss were 3$\times$10$^{-18}$ cm$^2$ and 7$\times$10$^{-19}$ 
cm$^2$, respectively. For $C_6H_6$-ion collisions at 50-110 MeV, the domination of 
direct ionization processes is likely to be even more pronounced. 

For the light-field experiments, determination of the peak field value is 
somewhat less difficult, but note the following facet that has hitherto not 
been widely acknowledged. Just as the ion-$C_6H_6$ interaction accesses 
a range of $b$-values ($b>$3 {\AA}), and, hence, exposes the molecules in the 
interaction zone to a corresponding range of applied fields, so in the laser 
case there is a spatial distribution of intensities that gives rise to a 
corresponding distribution of fields. Such 
spatial distribution of the laser-induced field, and the ion distribution 
that results from it, was measured using a method that has recently 
been described in detail \cite{spatial}. The spatial 
field distributions obtained in both the ion-impact and laser-based 
experiments are 
shown in Figs. 1 and 2, respectively. For convenience of comparison,  we  
express the field magnitude obtained in the ion-impact experiments 
in terms of an effective intensity that can be directly compared with an 
easily-determined parameter in the laser experiments, namely the light 
intensity. Judicious choice of operating conditions enable 
the spatial field distributions to be very similar in the two sets of experiments. 
The examples shown are for a 100-MeV beam of Si$^{8+}$ ions, readily accessed 
in tandem accelerators, and for a 100-fs long laser pulse of peak intensity 
2$\times$10$^{15}$ W cm$^{-2}$, accessed by focusing a 1 mJ beam of 806 nm 
wavelength from a Ti:sapphire laser to a spot size of $\sim$25 $\mu$m.  

Fig. 3 shows typical ionic fragmentation patterns obtained when $C_6H_6$ is 
irradiated by 100-MeV Si$^{3+}$ ions and 35-ps-long laser pulses of 
532 nm wavelength. The peak intensity in the latter case was 
$\sim$8$\times$10$^{13}$ W cm$^{-2}$ and, by using a small (2 mm) aperture 
at the entrance of our TOF spectrometer, the lower part of the intensity range 
that we accessed was $\sim$5$\times$10$^{12}$ W cm$^{-2}$.   In the case of 
the ion impact data, the peak value of effective intensity at an impact 
parameter of 3 {\AA} was $\sim$5$\times$10$^{14}$ W cm$^{-2}$. We also 
conducted experiments with Si$^{8+}$ and F$^{7+}$ ions, and with 100-fs duration 
laser pulses (of 806 nm wavelength), with peak intensities in the range 
10$^{13}$-5$\times$10$^{16}$ W cm$^{-2}$. The gross features of the 
measured ionic fragmentation patterns remained essentially 
unaltered under all these conditions, although the highest laser intensities 
gave rise to a larger degree of multiple ionization (that could also be 
directly correlated with ion beam data). Illustrative data are shown in Fig. 4, 
for 110 MeV Si$^{8+}$ impact and irradiation by 100fs laser pulses of intensity 
5$\times$10$^{15}$ W cm$^{-2}$. Here, 
a larger (15 mm) aperture was used at the entrance of our TOF spectrometer in
our laser experiments, thereby 
giving access to the entire intensity range shown in Fig. 2. 
Recent work \cite{spatial} has shown that the 
lower-intensity portions of this range dominate the overall dynamics because 
a very much larger proportion of the focal volume that is accessed by the 
spectrometer samples the lowest intensities. Under these circumstances, the 
contribution of the parent 
$C_6H_6^{~+}$ ion to the mas spectrum is very much enhanced (not shown in the 
figure). The central portion of the focal volume samples the peak intensity 
region and, consequently, multiple ionization events are also more in evidence in 
the spectrum shown in Fig. 4. A noteworthy feature is the somewhat unexpected 
preponderence of H$^+$ and H$_2^{~+}$ fragments, indicating that extremely 
energetic proceeses occur in this intensity regime that overcomes the intrinsic 
strength of the aromatic ring structure and causes extensive fragmentation, 
opening dissociation channels that are seldom accessed in electron impact  
experiments. In order to preclude the possiblity of diverting from the main 
focus of attention in this Letter, we refrain from discussing the details of 
the ionic fragmentation patterns  
(such discussion can be found in the context of $ns$ and $ps$  
laser-induced fragmentation of $C_6H_6$ in ref.\cite{previous}). We focus on the
fact that the data presented 
here clearly does show a surprising degree of similarlity in 
the morphology of the ion-induced and light-induced fragmentation 
patterns. 

So, what lessons are to be drawn from our attempts to study the correspondence 
between ion-induced and light-induced strong fields and their effect on molecular 
dynamics? Clearly, the first lesson is that the observed 
similarities in the mass spectra show that this correspondence is worthy of further 
pursuit, notwithstanding the ``obvious" differences in the nature of
the fields 
generated by non-relativistic charged particle beams and pure electromagnetic 
radiation. The second lesson originates in the practical consideration that 
irradiation of matter by either type of field entails exposure to a {\em range} 
of field intensities; single-valued fields are unlikely to be encountered, 
either in nature or in the laboratory. The spatial variation of field intensity 
gives rise to ``focal volume effects" amd makes it mandatory to ensure that 
equivalent intensity ranges are accessed in studies of field-matter interactions 
in order to make meaningful comparisons about the effects on the overall dynamics 
of the time duration of each type of field. We believe that we have succeeded in 
achieving this equivalence in our experiments. The somewhat surprising result to 
emerge is that, at least for $C_6H_6$, whether the intense field is of picosecond 
or attosecond duration is of little consequence as far as the overall 
ionic fragmentation pattern is concerned. Thirdly, the other major difference between 
the two types of fields concerns their directional properties: the laser-induced 
field has a well-defined direction as it originate in light that is 
linearly-polarized. But, the direction of the ion-induced field 
changes in the course of the collision process. Our data indicate that these  
properties seem unimportant to the overall dynamics. It 
may be argued that for ion-induced fields the change of direction is too 
fast to be of consequence. 

Can our results be interpreted within the framework 
of molecular quantum mechanics? One attempt has been made to study intense 
laser-induced ionic fragmentation within the framework of molecular orbital 
energies \cite{rottke}, but application of fields that last only for, say 30 
$as$, implies an energy uncertainty of $\sim$22 eV. This makes quantal 
treatments of the ionization dynamics with conventional molecular states 
irrelevant. 

Gratitude is expressed to colleagues who have contributed their mite to 
the accelerator-based and intense-laser experiments on molecules in the 
course of several years: F. A. Rajgara, V. R. Bhardwaj, G. Ravindra Kumar, 
U. T. Raheja, V. Krishnamurthi, C. Badrinathan, K. Vijayalakshmi, S. Banerjee, 
C. P. Safvan, M. Krishnamurthy, V. Kumarappan, E. Krishnakumar, K. Nagesha, 
and A. K. Sinha. The high-energy femtosecond laser was partially funded by 
the Department of Science and Technology. Some experiments were conducted 
at the Nuclear Science Centre, New Delhi.

\begin{figure}
\caption{The radial and axial distribution of applied field within the 
interaction volume for $Si^{8+}$-$C_6H_6$ collisions. The magnitude of 
the field is expressed in terms of an effective intensity in order to 
aid comparison with the equivalent distribution pattern obtained in 
laser-$C_6H_6$ interactions.}
\end{figure}

\begin{figure}
\caption{The spatial distribution of applied field within the focal 
volume for laser-$C_6H_6$ interactions (see text).}
\end{figure}

\begin{figure}
\caption{Ion fragmentation patterns obtained upon irradiation of 
$C_6H_6$ by a) 100-MeV Si$^{3+}$ ions (similar patterns were obtained 
using Si$^{8+}$ and F$^{7+}$ ions), and b) 35-ps-long laser pulses 
of 532 nm wavelength and peak intensity of 8$\times$10$^{13}$ 
W cm$^{-2}$.}
\end{figure}

\begin{figure}
\caption{Fragmentation of $C_6H_6$ by a) 110-MeV Si$^{8+}$ ions and 
b) 100-fs-long laser pulses of 806 nm wavelength and peak intensity 
of 5$\times$10$^{15}$ W cm$^{-2}$.}
\end{figure}


\begin{references}

\bibitem{ww} J. D. Jackson, in {\it Classical Electrodynamics} (Wiley, 
New York, 1962), Chap. 15.
\bibitem{moshammer}S. Keller {\it et al.}, Phys. Rev. A {\bf 55}, 4215 (1997), 
R. Moshammer {\it et al.}, Phys. Rev. Lett. {\bf 79}, 3621 (1997). 
\bibitem{rhodes}K. Boyer {\it et al.}, IEEE Trans. Plasma Sci. {\bf 16}, 541 
(1988), T. S. Luk {\it et al.}, Phys. Rev. A {\bf 48}, 1359 (1993).
\bibitem{nsc}V. R. Bhardwaj {\it et al.}, Phys. Rev. A {\bf 58}, 3849 (1998), 
{\it ibid.} {\bf 59}, 3105 (1999).
\bibitem{ch4}D. Mathur {\it et al.}, Phys. Rev. A  {\bf 50}, R7 (1994),
J. Phys. B {\bf 27}, L603 (1994).
\bibitem{tonuma}T. Tonuma {\it et al.}, J. Phys. B {\bf 17}, L317 (1984).
\bibitem{schlachter}A. S. Schlachter {\it et al.}, Phys. Scr. {\it T3}, 143 
(1983).
\bibitem{spatial}S. Banerjee {\it et al.}, 
J. Phys. B {\bf 32}, 4277 (1999), Phys. Rev. A {\bf 60}, R3369 (1999).
\bibitem{previous}V. R. Bhardwaj {\it et al.}, Phys. Rev. A {\bf 59}, 1392 
(1999), H. J. Neusser {\it et al.}, Int. J. Mass Spectrom. Ion Processes 
{\bf 60}, 147 (1984),
and references therein.
\bibitem{rottke}H. Rottke {\it et al.}, J. Phys. B {\bf 31}, 1083 (1998).


\end{references}
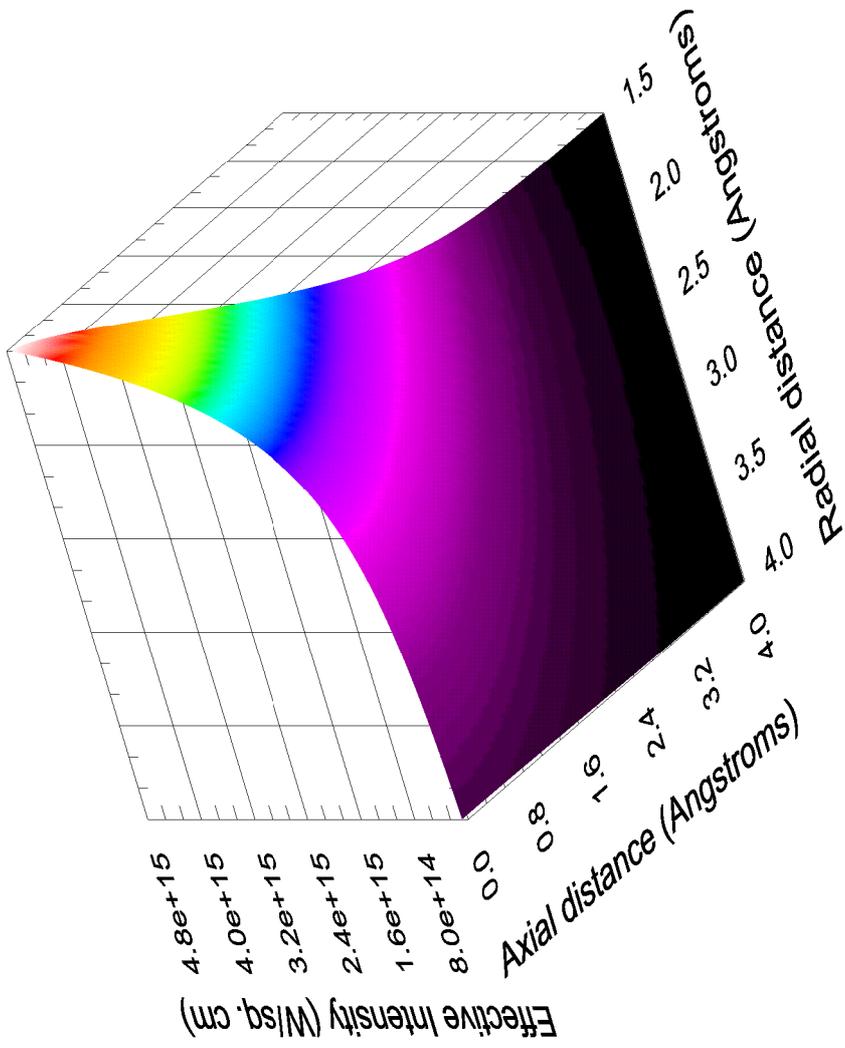
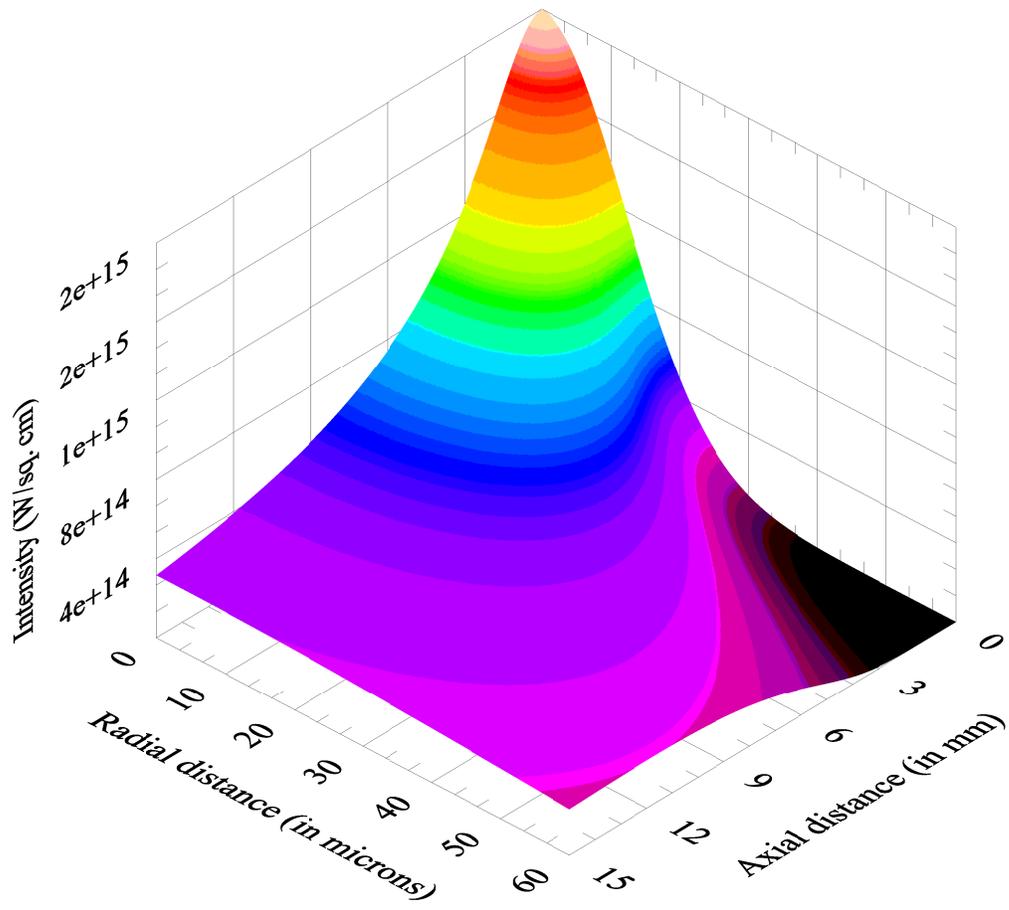
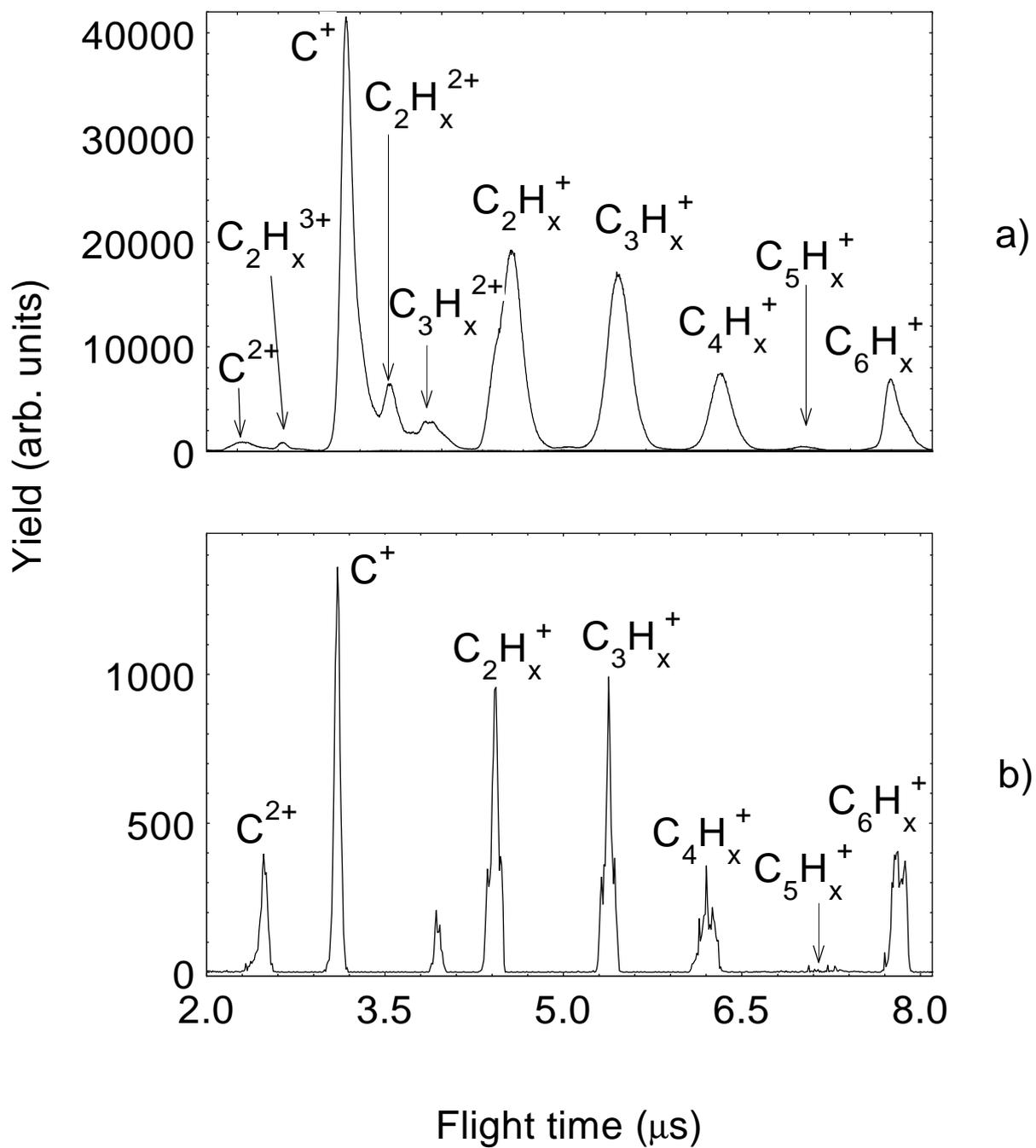
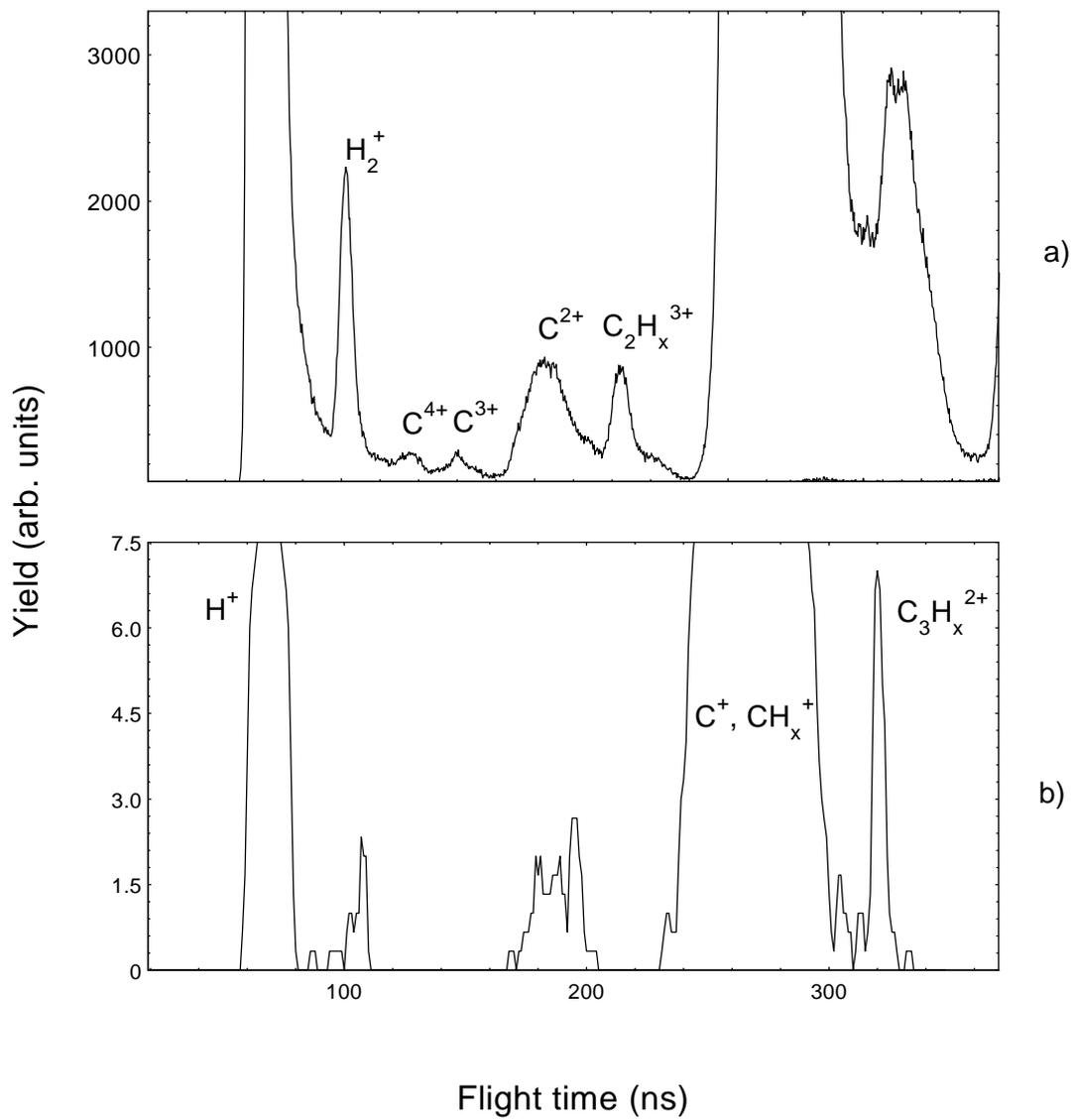
\end{document}